\documentclass{PoS}

\def\simleq{\; \raise0.3ex\hbox{$<$\kern-0.75em \raise-1.1ex\hbox{$\sim$}}\; }
\def\simgeq{\; \raise0.3ex\hbox{$>$\kern-0.75em \raise-1.1ex\hbox{$\sim$}}\; }

\newcommand{\GeV}{{\rm GeV}}
\newcommand{\TeV}{{\rm TeV}}
\newcommand{\PeV}{{\rm PeV}}

\newcommand{\kpc}{{\rm kpc}}

\newcommand{\cm}{{\rm cm}}
\newcommand{\km}{{\rm km}}

\newcommand{\s}{{\rm s}}
\newcommand{\sr}{{\rm sr}}

\usepackage{subfigure}

\title{Gamma-ray and neutrino diffuse emissions of the Galaxy above the TeV}

\ShortTitle{Gamma-ray and Neutrino Galactic emissions above the TeV}

\author{Daniele Gaggero\\
        {SISSA and INFN, via Bonomea 265, I-34136 Trieste, Italy}\\
        E-mail: \email{daniele.gaggero@sissa.it}}
        
\author{\speaker{Dario Grasso}\thanks{A footnote may follow.}\\
        INFN and Dipartimento di Fisica ``E. Fermi", Pisa University, Largo B. Pontecorvo 3, I-56127 Pisa, Italy\\
        E-mail: \email{dario.grasso@pi.infn.it}}
        
\author{Antonio Marinelli\\
        INFN and Dipartimento di Fisica ``E. Fermi", Pisa University, Largo B. Pontecorvo 3, I-56127 Pisa, Italy\\
        E-mail: \email{antonio.marinelli@pi.infn.it}}        

\author{Alfredo Urbano\\
        {SISSA and INFN, via Bonomea 265, I-34136 Trieste, Italy}\\
        E-mail: \email{alfredo.urbano@sissa.it}} 

\author{Mauro Valli\\
        {SISSA and INFN, via Bonomea 265, I-34136 Trieste, Italy}\\
        E-mail: \email{mauro.valli@sissa.it}}

\abstract{
As recently shown, Fermi-LAT measurements of the diffuse gamma-ray emission from the Galaxy favor the presence of a smooth softening in the primary cosmic-ray spectrum with increasing Galactocentric distance. 
This result can be interpreted in terms of a spatial-dependent rigidity scaling of the diffusion coefficient.  
The {\tt DRAGON} code was used to build a model based on such feature. 
That scenario correctly reproduces the latest Fermi-LAT results as well as local cosmic-ray measurements from PAMELA, AMS-02 and CREAM.
Here we show that the model, if extrapolated at larger energies, grasps both the gamma-ray flux measured by MILAGRO at 15 TeV and the H.E.S.S. data from the Galactic ridge, assuming that the cosmic-ray spectral hardening found by those experiments at about 250 GeV/n is present in the whole inner Galactic plane region.
Moreover, we show as that model also predicts a neutrino emission which may account for a significant fraction, as well as for the correct spectral shape, of the astrophysical flux measured by IceCube above 25 TeV.  
}

\FullConference{The 34th International Cosmic Ray Conference,\\
		30 July- 6 August, 2015\\
		The Hague, The Netherlands}

\begin{document}

\section{Introduction}

Cosmic-ray (CR) transport in the Galaxy and secondary $\gamma$-ray diffuse emission are generally modeled assuming that charged particle species undergo a statistically homogeneous and isotropic diffusion due to their scattering onto the Galactic magnetic field random fluctuations. This amounts to treat the diffusion coefficient as a spatially independent scalar function of the particle rigidity. Although this simplified approach does not stand on solid theoretical grounds -- the random magnetic field power is not expected to be homogenous and the presence of a regular field component should break the isotropy -- nevertheless it provides a quite successful description of a wide number of experimental data. 
Indeed, CR spectra of primary and secondary CR species can accurately be reproduced in that {\it conventional} framework at least up to few hundred GeV/nucleon (see {\it e.g.} \cite{Evoli:2015vaa} for some recent updated results). 
 
It should be noted, however, that charged CR species provide a rather local probe of the Galactic environments. Locally detected secondary species, as the boron, which are generally used to tune the relevant propagation parameters, are produced within few kpc from the Earth, therefore they do not probe the conditions in the central region of the Galaxy. 
Instead, secondary $\gamma$-rays, produced by the interaction of CR nuclei (electrons) with the interstellar gas (radiation), offer a deeper insight of the Galaxy.  
Interestingly, although the {\it conventional} picture implemented with numerical codes like {\tt GALPROP} \cite{Galpropweb,Moskalenko2002} reproduces the main features of the Galactic diffuse emission, several anomalies suggest that some main ingredients may be missing in that approach.   

We start noticing that {\it conventional} models cannot explain the large $\gamma$-ray flux measured by the Milagro observatory from the inner GP region ($|b| < 2^\circ$, $30^\circ < l < 65^\circ$) at $15~\TeV$ median energy \cite{Abdo:2008if}.  
An {\it optimized} model which was proposed \cite{Strong:2004de} to account for the EGRET GeV excess and came out to reproduce Milagro result as well, was subsequently excluded by Fermi-LAT~\cite{Abdo:2009mr}. 
This problem holds also for the most updated conventional models tuned to reproduce updated Fermi-LAT data \cite{FermiLAT:2012aa} (see below). 
Another intriguing discrepancy was found between the predictions of conventional models and high-energy $\gamma$-ray H.E.S.S. data for the Galactic ridge region ($| l | < 0.8^\circ$,  $| b | < 0.3^\circ$) \cite{Aharonian:2006au}.
The diffuse spectrum measured by H.E.S.S. for $0.3 \simleq E \simleq 10~\TeV$ in that region is well described by a power law with index $2.29 \pm 0.07~ \pm 0.20$. This is significantly harder than expected from $\pi_0$-decay if a CR spectral shape identical to that found in the solar neighborhood is assumed in the GC region. 
While in \cite{Aharonian:2006au} the authors suggest such discrepancy to be originated by the proximity between the dense molecular gas with a CR source in that region, no compelling evidence supporting that interpretation has been provided and other explanations worth to be considered.      

It is important to point out that several troubles for conventional models arise also at lower energies: {\tt GALPROP} based conventional models tuned to reproduce the diffuse $\gamma$-ray over most of the sky 
(see e.g. \cite{Galpropweb,Moskalenko2002} and references therein) systematically underestimate the measured flux in the inner Galactic Plane (GP) region above few GeV.
This problem, which was pointed-out by the Fermi collaboration itself \cite{FermiLAT:2012aa}, has been independently confirmed using updated Fermi data and point source catalogue (3FGL) \cite{Gaggero:2014xla}.

Other independent hints of the inadequacy of the standard CR propagation scenario are coming also from the newly born high-energy neutrino astronomy. 
In the last few years the IceCube collaboration claimed the detection of 37 neutrino events above  $30~\TeV$ corresponding to an excess with respect to the atmospheric background of $5.7~\sigma$ \cite{Aartsen:2014gkd}. 
The inferred flavor composition is compatible with a mixture of electronic, muonic and tauonic neutrino in equal amounts as expected if their origin were astrophysical. 
A recent analysis \cite{Aartsen:2014muf} of 641 selected events with vertices contained in the detector allowed the IceCube collaboration to lower the energy threshold and to measure the extraterrestrial diffuse neutrino spectrum to be  $\Phi_\nu = 2.06_{- 0.3}^{+0.4} \times 10^{-18}~\left(E_\nu/10^5~\GeV\right)^{-2.46 \pm 0.12}~\GeV^{-1}\cm^{-2}\sr^{-1}\s^{-1}$ for $25~\TeV <  E_\nu < 1.4~\PeV$. 
This spectrum is significantly softer than expected from most extragalactic sources (e.g. AGNs or $\gamma$-ray bursts). This fact, together with a predominancy of down-going events and their concentration - though still not statistically significant - toward the GC region, suggests the presence of a significant Galactic component of the detected astrophysical neutrino flux \cite{Spurio:2014una}. 
%
Although the Galaxy is known to be a guaranteed source of a neutrinos originated from the hadronic interactions between the Galactic CR sea and the diffuse interstellar gas, {\it conventional} models  predict an emission which is considerably smaller and softer than the flux measured by IceCube (see {\it e.g.} \cite{Evoli:2007iy} and ref.s therein). 

In this contribution we discuss a possible common solution of all those anomalies. 
This is based on a scenario, recently presented in \cite{Gaggero:2014xla}, featuring a radial dependence for both the rigidity scaling index $\delta$ of the diffusion coefficient and the convective wind. Since the model assumes $\delta$ {increasing} with $R$ and it is tuned to reproduce CR locally observed spectra, 
it predicts a hardening of CR propagated spectrum, hence of the secondary $\gamma$-ray emission, in the inner Galaxy. The model also accounts for the CR proton and Helium spectral harding inferred from PAMELA \cite{Adriani:2011cu}, CREAM \cite{Ahn:2010gv} and AMS-02 \cite{Aguilar:2015ooa} at few hundred GeV/n.  We will see as such a feature is actually required to get a consistent description of Fermi-LAT and upper-TeV $\gamma$-ray data by Milagro and H.E.S.S. 

\section{The KRA$_\gamma$ setup}

We treat CR transport in the Galaxy by means of the {\tt DRAGON} code \cite{Evoli:2008dv,Dragonweb}. 
Similarly to other numerical CR transport codes (e.g. {\tt GALPROP} \cite{Galpropweb,Moskalenko2002}), {\tt DRAGON} solves numerically the diffusion equation for a given source term and null boundary conditions away from the Galactic disk.  
While {\tt DRAGON} shares with {\tt GALPROP}  the public CR spallation cross sections and the interstellar gas and radiation distributions (other choices of those quantities are however possible in {\tt DRAGON}), differently from {\tt GALPROP} public versions {\tt DRAGON} allows, among several other innovative features, to adopt a spatial dependent diffusion coefficient and convective velocity.
This feature has been used to implement the KRA$_\gamma$ models discussed in this contribution (see also \cite{Gaggero:2014xla,Gaggero:2015xza}). 
Those models assume that the exponent $\delta$ determining the rigidity dependence of the CR diffusion coefficient
\begin{equation}
D(R,z,\rho) =  D_0 \left(\rho/\rho_0\right)^{\delta(R)}\ \exp(z/z_t)
\end{equation}
has the following Galactocentric radial dependence: $\delta(R) = A R + B$ where $A = 0.035~\kpc^{-1}$ and $B = 0.21$ so that $\delta(R_{\odot}) = 0.5$. 
Concerning the vertical dependence of the diffusion coefficient we assume $z_t = 4~\kpc$  (we checked that our results do not change significantly considering larger values of $z_t$). 
As we mentioned,  the models also adopt a convective wind for $R < 6.5~\kpc$ with velocity $V_C(z) {\hat z}$ ($z$ is the distance from the GP) vanishing at $z = 0$ and growing as $dV_c/dz = 100~\km~ \s^{-1}~ \kpc^{-1}$.
%
Those model parameters have chosen to consistently reproduce the slope and angular distribution of the diffuse $\gamma$-ray emission of the Galaxy measured by Fermi-LAT as well as the main local CR observables (proton and Helium spectra and the B/C ratio) \cite{Gaggero:2014xla}.

\begin{figure}[tbp]
\centering
\includegraphics[width=0.45\textwidth]{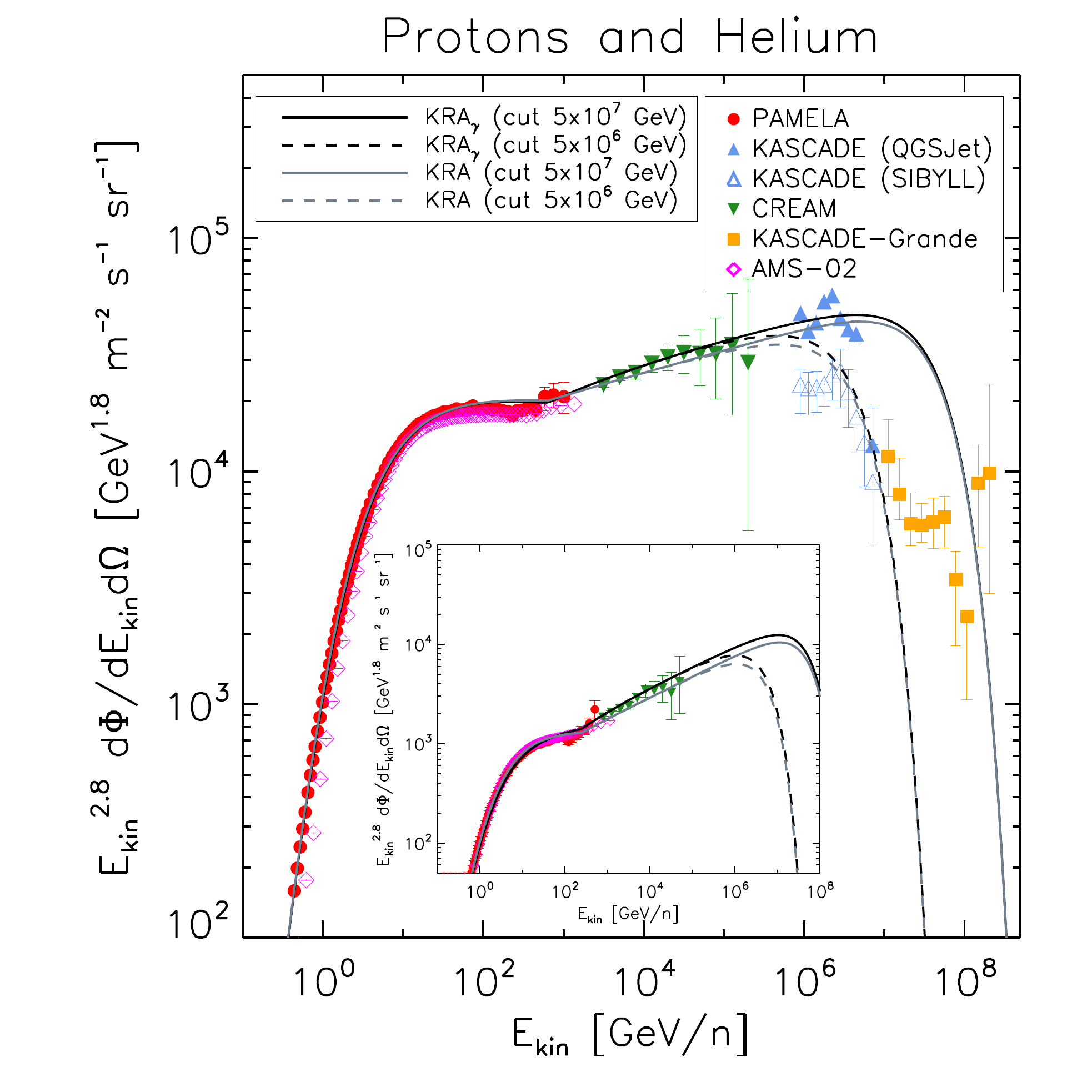}
\caption{The computed proton and Helium (in the insert) spectra at the solar circle are compared with a representative set of experimental data. All spectra account for solar modulation with a potential $\Phi = 0.5~{\rm GV}$. }
\label{fig:proton_He}
\end{figure}

Since Fermi-LAT measured the spectrum of the diffuse Galactic emission up to few hundred GeV, some further hypothesis has to be done about the spectrum of primary CR above the TeV.  
In particular, concerning the $p$ and He spectral hardening inferred from PAMELA \cite{Adriani:2011cu} - recently confirmed by AMS-02 \cite{Aguilar:2015ooa} -
and CREAM \cite{Ahn:2010gv} data above $\sim 250~\GeV/{\rm n}$, we consider two cases: 1) KRA$_\gamma$ {\it no hard.} model:  the hardening is assumed to be a local effect - {\it e.g.} originated by nearby supernova remnants - which averages out on large scales hence introducing no feature in the Galactic CR population. 
 2) KRA$_\gamma$ model: in this case the hardening is taken to be representative of the whole Galactic population. 
In both cases we assume that above 250 GeV/n the CR source spectra extend steadily up to an exponential cutoff varying in the range $5 < E_{\rm cut}/Z < 50~{\rm PeV}$ to roughly agree with KASCADE \cite{Antoni:2005wq} and KASCADE-Grande data \cite{Apel:2013dga} (see fig.\,\ref{fig:proton_He}). 

\section{The diffuse $\gamma$-ray spectrum above the TeV}

It was shown in \cite{Gaggero:2014xla}, that both the KRA$_{\gamma}$ {\it no hard.} and KRA$_{\gamma}$ models provide good fits of the $\gamma$-ray diffuse emission measured by Fermi-LAT all over the sky, in particular towards the inner GP region.  In \cite{Gaggero:2015xza} we extended that computation above the TeV. 
Starting from the propagated proton and Helium distributions we computed the hadronic emission integrating the expression of the $\gamma$-ray emissivity \cite{Kamae:2006bf} along the line-of-sight.
%
%
For the gas components we used the distributions taken from the latest public version of the {\tt GALPROP} package \cite{Galpropweb,FermiLAT:2012aa}.

\begin{figure}[tbp]
\centering
\includegraphics[width=0.45\textwidth]{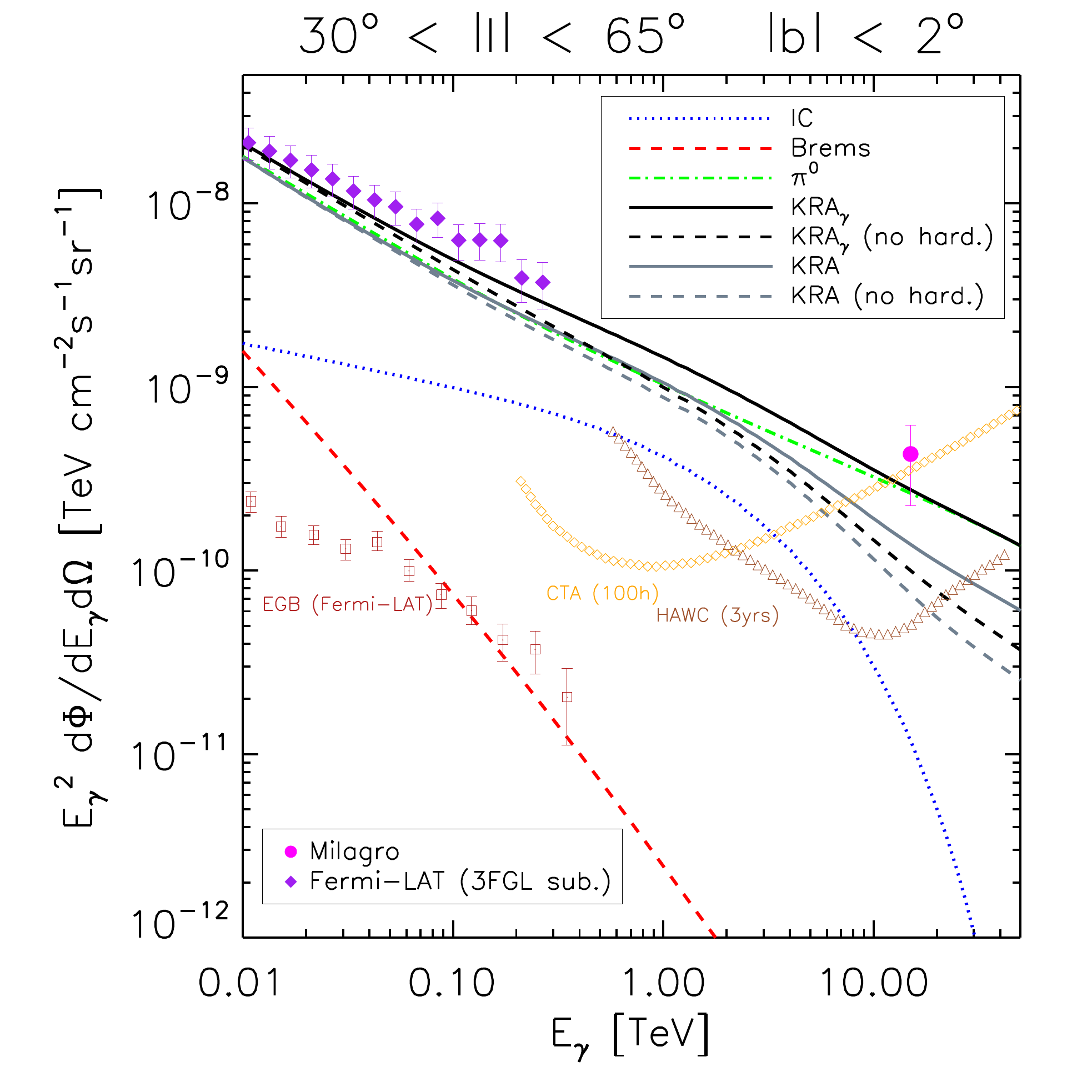}
\includegraphics[width=0.45\textwidth]{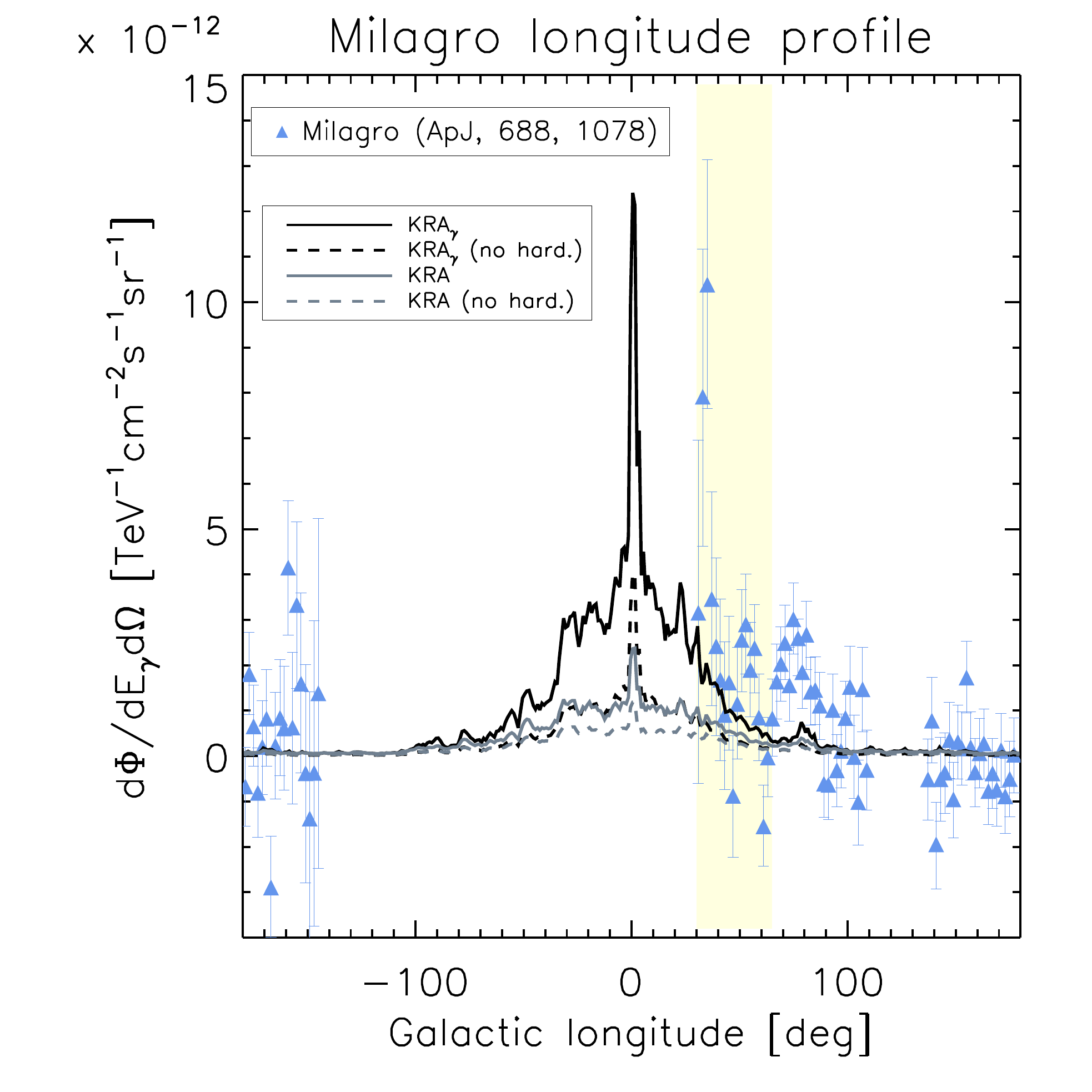}
\caption{The diffuse emission $\gamma$-ray spectrum (left panel) from the inner
Galactic plane ($|b| < 2^\circ$,   $30^\circ < l < 65^\circ$) computed for the reference models considered in this Letter is compared with
Fermi-LAT and Milagro data. The expected sensitivity of HAWC and CTA  are reported. The spectral components are shown for the KRA$_\gamma$ model only.
In the right panel the longitude profile at 15 TeV for the same models is compared with HAWC data.} 
\label{fig:milagro}
\end{figure}

In Fig. \ref{fig:milagro} we compare our predictions for the KRA$_\gamma$ and KRA$_{\gamma}$ {\it no hard.} with Fermi-LAT and Milagro diffuse emission data from the same inner GP region. In the same plot we also report {\tt DRAGON} predictions for a representative conventional model (KRA) - again with and without hardening - sharing the same main properties of the KRA$_\gamma$ model at the solar circle ({\it i.e.} $\delta = 0.5$ and $dV_c/dz =  0$  in the whole Galaxy).

 The reader can see as conventional models, which under-predict Fermi-LAT data in agreement with \cite{Gaggero:2014xla}, more seriously fail reproducing Milagro result.
While the presence of hardening in the whole CR Galactic population clearly enhances the $\gamma$-ray flux above few hundred GeV, this is not enough to account for the flux measured by Milagro from the inner GP at 15 TeV.  
The KRA$_\gamma$ setup, instead, is more successful -- especially if the CR hardening is assumed. 
In the right panel of Fig.\ref{fig:milagro} we also compare our results with the longitude profile at 15 TeV measured by Milagro. 
Although the data do not allow any strong claim, we notice a better fit of the $KRA_{\gamma}$ model, although this should only be taken as a hint.

\begin{figure}[tbp]
\centering
\includegraphics[width=0.45\textwidth]{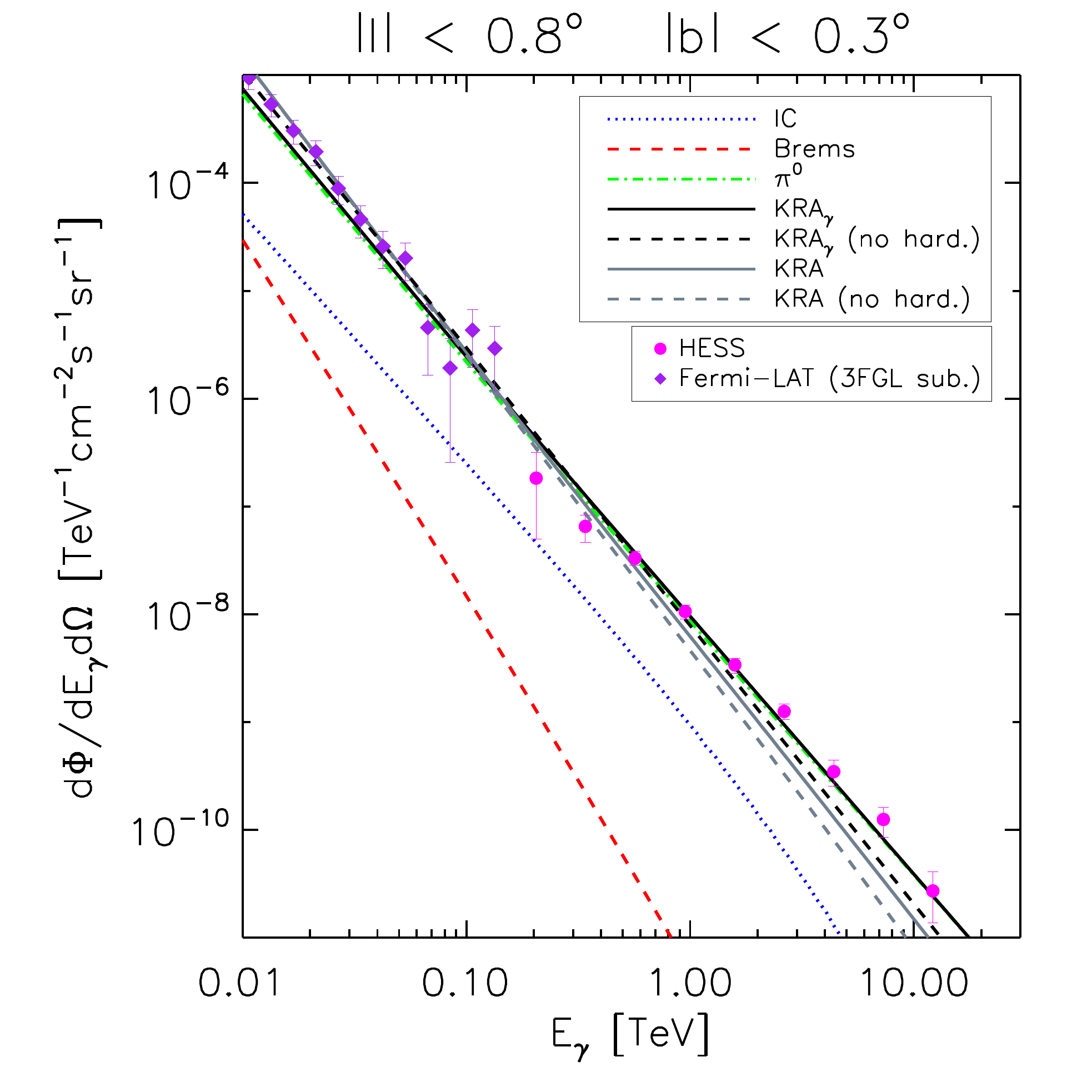}
\caption{The computed $\gamma$-ray diffuse emission from the Galactic ridge region is compared with Fermi-LAT and H.E.S.S. data. For each model the spectrum normalization has been varied to minimize the $\chi^2$ against the data. The spectral components are shown for the KRA$_\gamma$ model only. 
}
\label{fig:hess}
\end{figure}

Exploring a region not covered by Milagro, the H.E.S.S. observation of the Galactic ridge offers a further test of our model.  Interestingly, also for that region the model provides a good fit of Fermi-LAT and H.E.S.S. results (see Fig.\,\ref{fig:hess}).  
We also cross-check adopting a more realistic gas distribution \cite{Ferriere:2007yq} in the inner Galaxy, and we rescale the models by a factor of $0.3$ to minimize the $\chi^2$ against the data: This factor is justified by the smaller value of the conversion factor ($X_{\rm CO}$) between the ${\rm H_2}$ column density and the CO line brightness temperature in the central region~\cite{Ferriere:2007yq}.

\section{Implications for the Galactic neutrino emission}

The presence of a hard CR spectrum in the inner Galactic plane region as predicted by the KRA$_{\gamma}$ setup has relevant consequences also for the diffuse neutrino emission produced by CR hadron scattering onto the interstellar gas. 
In order to estimate such an effect, we computed the $\nu_e$ and $\nu_\mu$ production spectra similarly to what done for $\gamma$-rays. For both flavors we used the emissivities provided in \cite{Kamae:2006bf} for projectile energies below $\sim 500$ TeV, while we adopted the ones provided in \cite{Kelner:2006tc} above that range.
We also accounted for neutrino oscillations which redistribute the composition almost equally among all the three flavors.  
We only considered proton and Helium CRs/gas -- as done for $\gamma$-rays -- since heavier nuclear species give a negligible contribution in the energy range we cover in this work. 

\begin{figure}[tbp]
\centering
\includegraphics[width=0.45\textwidth]{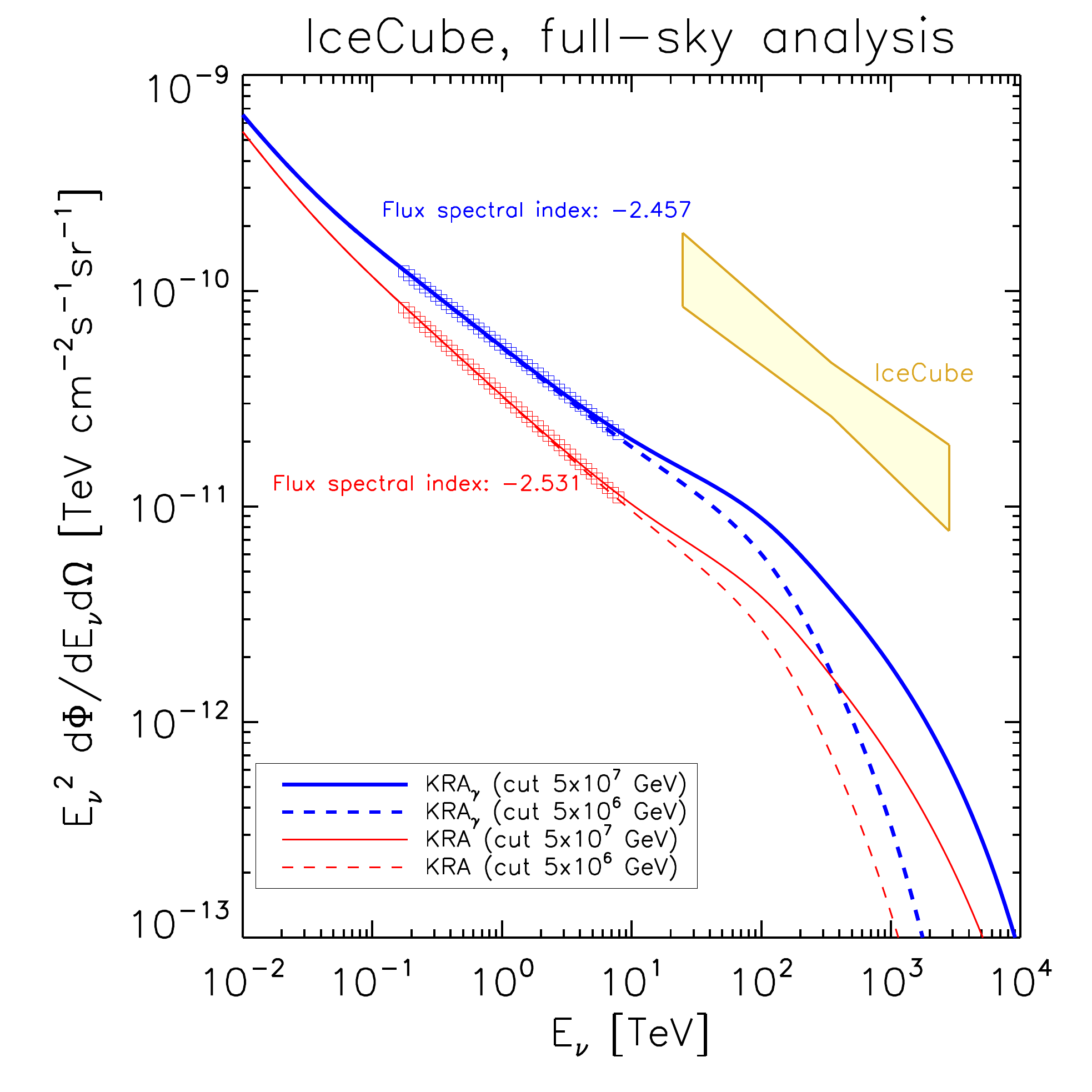}
\caption{The full-sky neutrino spectrum computed for the KRA$_\gamma$ and KRA models (both with global CR hardening) and two choices of the CR high-energy cutoff is compared with IceCube result \cite{Aartsen:2014muf}.}
\label{fig:nu_spectra}
\end{figure}

In Fig.\,\ref{fig:nu_spectra}  we show the total neutrino spectrum (all flavors, including antiparticles) computed for the KRA$_\gamma$ (blue lines) in the presence of a CR spectral hardening in the whole Galaxy. 
In order to account for the uncertainties on the high energy tail of the primary CR spectra we considered two values of the CR source spectrum cutoff $E_{\rm cut} = 5\; \PeV$ (dashed lines) and
$50\; \PeV$ (solid lines) roughly bracketing KASCADE and KASCADE-Grande data (see Fig.\,\ref{fig:proton_He}).
For comparison we also show our predictions for the convetional KRA model for the same values of  $E_{\rm cut}$ (red  lines). 
In all cases we assume the CR hardening at $\sim 250$ GV to be present in the whole Galaxy. 
The reader can see from that figure that, while the KRA setup predicts an undetectable flux -- in agreement  with previous results for other conventional models \cite{Evoli:2008dv} -- the KRA$_\gamma$ model is instead able to explain a significant fraction of the flux measured by IceCube in terms of Galactic diffuse emission. Interestingly, for that model, the flux spectrum below 100 TeV - where the Galactic contribution is expected to be larger - almost follows a power law spectrum with an index which is very close to that measured by IceCube.  

Setting a threshold energy at $25$ TeV, with a cosmic-ray source cutoff $E_{\rm cut} \,=\, 50$ PeV we obtain an expected event rate representing 40\% of the complete sample of 37 events reported, well above the expected rate of atmospheric muons and atmospheric neutrinos after the veto conditions imposed by the IceCube collaboration. On the other hand, considering a threshold energy of $60$ TeV the expected event rate is about 30\% of the corresponding measured sample of 20 events. Decreasing the $E_{\rm cut}$ to $5$ PeV the expected event rates become 20\% (above $25~\TeV$) and 10\% (above $60~\TeV$) of the measured IceCube samples. 
This result could soon be confirmed for the southern hemisphere emission by ANTARES experiment \cite{Aguilar:2010ab,Marinelli_talk} selecting a low energy sample of events ($<\,60$ TeV) and smaller portion of the sky, and accurately verified when IceCube and the incoming KM3NeT \cite{Adrian-Martinez:2012qpa} obtain a large catalog of astrophysical neutrino events. 

Concerning the angular distribution of IceCube events, we notice that -- although IceCube collaboration did not report any statistically significant anisotropy yet -- a recent analysis \cite{Ahlers:2015moa} showed as a Galactic component producing up to $50\%$ of the astrophysical neutrino flux measured by IceCube is compatible with the angular distribution of high-energy events detected by that experiment.  Therefore, our model is consistent with those results.
Interestingly, the IceCube collaboration just published a new analysis \cite{Aartsen:2015ita} pointing to different flux normalizations and spectral indexes in the North and South hemisphere which may be interpreted with the presence of a component from the inner Galaxy as suggested in the above.

\section{Conclusions}

Conventional CR transport models - assuming uniform diffusion/advection properties - provide unsatisfactory description of the diffuse Galactic $\gamma$-ray diffuse emission measured by Fermi-LAT toward the inner Galactic plane. 
Moreover these models cannot reproduce measurements taken in the same region above the TeV. 
We showed that a recently proposed scenario assuming a proper radial dependence for both the rigidity scaling index $\delta$ of the diffusion coefficient and the convective wind, which allows to reproduce Fermi-LAT results \cite{Gaggero:2014xla}, also matches Milagro observations at 15 GeV.  Using a detailed gas model for the Galactic center regions, we showed as that model also consistently reproduces Fermi-LAT and H.E.S.S. measurements in the Galactic ridge region. 
Those results requires the CR spectral hardening found by PAMELA, AMS-02 and CREAM experiments at about 250 GV to be not a local effect but to be present also in inner Galactic plane region. 
We also considered the implications of our scenario for the neutrino diffuse emission of the Galaxy, Interestingly, we found that depending on the details of the CR spectra in the knee region, our model predicts that a fraction between $10~\%$ and $40\%$ of the signal measured by IceCube may be originated by CR scattering with the Galactic interstellar medium.    


\end{document}